\documentclass[preprint,12pt]{elsarticle}

\usepackage{subfig}
\usepackage{hyperref}
\usepackage{placeins}
\usepackage{amssymb}
\usepackage{amsmath}
\usepackage{comment}

\journal{Nuclear Instruments and Methods in Physics Research Section A Accelerators, Spectrometers, Detectors, and Associated Equipment}

\begin{document}

\begin{frontmatter}

\title{Development and Benchmarking of JANGOFETT: A Novel Geant4-Operated Fission Event Tracking Tool}
\author{\texorpdfstring{Liam Walker\textsuperscript{*}, Jack Shire\textsuperscript{*}, Jacob Jaffe, Payton Sprando, Jack Olinger, Alexander Chemey\textsuperscript{**}}}
\affiliation{organization={Oregon State University, Department of Nuclear Science and Engineering}, addressline={Oregon State University}, city={Corvallis}, postcode={97331}, state={OR}, country={United States}}

\tnotetext[equal]{These authors contributed equally to this work.} 
\tnotetext[corresp]{Corresponding author.\ E-mail address: \href{mailto:chemeya@oregonstate.edu}{chemeya@oregonstate.edu}.}

\begin{abstract}

Experiments measuring fission observables encounter false coincidences arising from timing overlap of separate fission product decays. Simulations of both fission observables and particles in detector systems exist, but have not yet been combined to produce accurate event-by-event outputs in a time-dependent manner. Geant4 is a powerful simulation tool for nuclear physics studies, but it does not typically handle multiple initial particles in a single simulation instance, nor does it feature high fidelity fission sampling. \textbf{\underline{J}ANGOFETT: \underline{A} \underline{N}ovel \underline{G}eant4-\underline{O}perated \underline{F}ission \underline{E}vent \underline{T}racking \underline{T}ool} has been developed to address this challenge. The tool utilizes simulated fission data from an external program in conjunction with Geant4, which has been modified to produce a single timeline of events over an entire simulated experiment. The physical accuracy of the simulated overlapping energy depositions within detectors has been verified via simulation of fission products from the spontaneous fission of $^{252}$Cf.

\end{abstract}


\end{frontmatter} 


\section{Introduction}
\label{intro}

There is significant interest in simulating fission and its subsequent detection in experimental apparatus \cite{randrup_calculation_2009}. Numerous experimental detection systems have been developed to correlate fission data at accelerator facilities around the world, with instruments such as DANCE (Detector for Advanced Neutron Capture Experiments) at the Los Alamos Neutron Science Center \cite{ullmann_dance_2004}, GAMMASPHERE at the Argonne Tandem Linac Accelerator System (ATLAS) \cite{lee_gammasphere_1990}, the $\nu$-ball-$\gamma$-spectrometer at the ALTO research platform \cite{lebois_nuball_2020}, SOPHIA (Study On FIssion with Aladin) at the GSI Helmholtz Centre \cite{boutoux_sofia_2013}, and the Superconducting Radioactive Isotope Beam Separator (BigRIPS) at RIKEN's Radioactive Isotope Beam Factory (RIBF) \cite{kubo_bigrips_2003}. Extensive simulations typically precede experiments with complex radiation detection systems to evaluate their potential and inform experiment sensitivity. This is often regarded as a necessary benchmark before any experiments occur, and it provides a valuable baseline for analysis code development.

There exist excellent software packages to facilitate Monte Carlo simulations assessing nuclear reactions and particle interactions within detectors such as MCNP \cite{los_alamos_scientific_laboratory_group_x-6_mcnp_1979} and Geant4 \cite{agostinelli_geant4simulation_2003}. MCNP excels in handling fission trends over whole populations by using an averaged fission model \cite{verbeke_fission_2015}. There are tools to simulate detector response with high fidelity within MCNP, such as MCNPX-Polimi \cite{clarke_mcnpx_polimi_2013}, but simulating multi-detector coincidence apparatus with charged particles remains a challenge. Geant4, developed by CERN, specializes in the simulated tracking of discrete simulated particles through matter, also using Monte Carlo methods. Utilizing object-oriented programming, the toolkit manages individual particles, de-excitations and material interactions within defined geometries \cite{j_allison_geant4_2006}\cite{allison_recent_2016}. Geant4 has superior ability to handle angular correlations, accurate nuclear de-excitations, and detector responses modeled on an event-by-event basis than alternatives. This enables algorithm development that can replicate experimental data with high fidelity.   

One notable weakness of Geant4 is found in the physics libraries for sampling fission fragments. Covariant trends in mass, energy, momenta, etc resulting from fission are not well represented. Geant4 normally handles nuclei (and daughter particles) sequentially, while fission is inherently multi-nuclear. There are tools that can produce realistic samples of fission physics, but these tools have varying degrees of implementation in Geant4 \cite{hecht_comparison_2014}. The models for fission are often outside of the base classes and are limited in their abilities to fully conserve energy, have incomplete validation, or lose information on important observables. Previous fission modifications have prioritized the correction of inaccuracies in the default fission models, \cite {constantin_simulation_2016} improving the simulation of isotopes, photons, and neutrons for use in statistical analysis of individual events \cite{tan_geant4_2017}. 

Geant4 resets the simulation world each time a new parent particle is introduced. Following fission, the resultant fission fragments undergo complex chains of de-excitation and decay, while new fission events continue to occur. This leads to the temporal overlapping of observables from both prior and newly initiated fission events, complicating data collection with false coincidences. By shifting the fission events with a time appropriate for a user-selected fission rate, Geant4 gains the ability to create a timeline of detector responses. This includes the effects of background radiation from delayed decays and their influence on fission product identification. 

There are many well-validated fission physics simulation tools useful for generation of covariant fission observables, such as FREYA \cite{verbeke_freya_2015} GEF \cite{schmidt_gef_2016}, TALYS \cite{fujio_talys_2023}, and CGMF \cite{talou_fission_2021}. These external fission codes can sample each fission event with accurate models, but are not designed to track the motion of the correlated fission products through space or their interactions with radiation detectors. The authors suggest that the strengths of such tools can be paired with the strengths of Geant4, providing simulated time correlated detector responses. 

 The work herein utilizes CGMF, a computational fission physics tool published by LANL, that includes all the most relevant physics observables produced in fission. \textbf{\underline{J}ANGOFETT: \underline{A} \underline{N}ovel \underline{G}eant4-\underline{O}perated \underline{F}ission \underline{E}vent \underline{T}racking \underline{T}ool} has been developed to enable Geant4 simulations that incorporate covariant time correlated observables important for experimental work. This work presents a general overview of JANGOFETT, the implementation of CGMF outputs into Geant4, and a verification study conducted with $^{252}$Cf(sf) using JANGOFETT to indicate multiple overlapping events occur in the simulated "experimental time".


\section{JANGOFETT Simulation Process}
\begin{figure}[ht]
\centering
\includegraphics[width=0.7\textwidth]{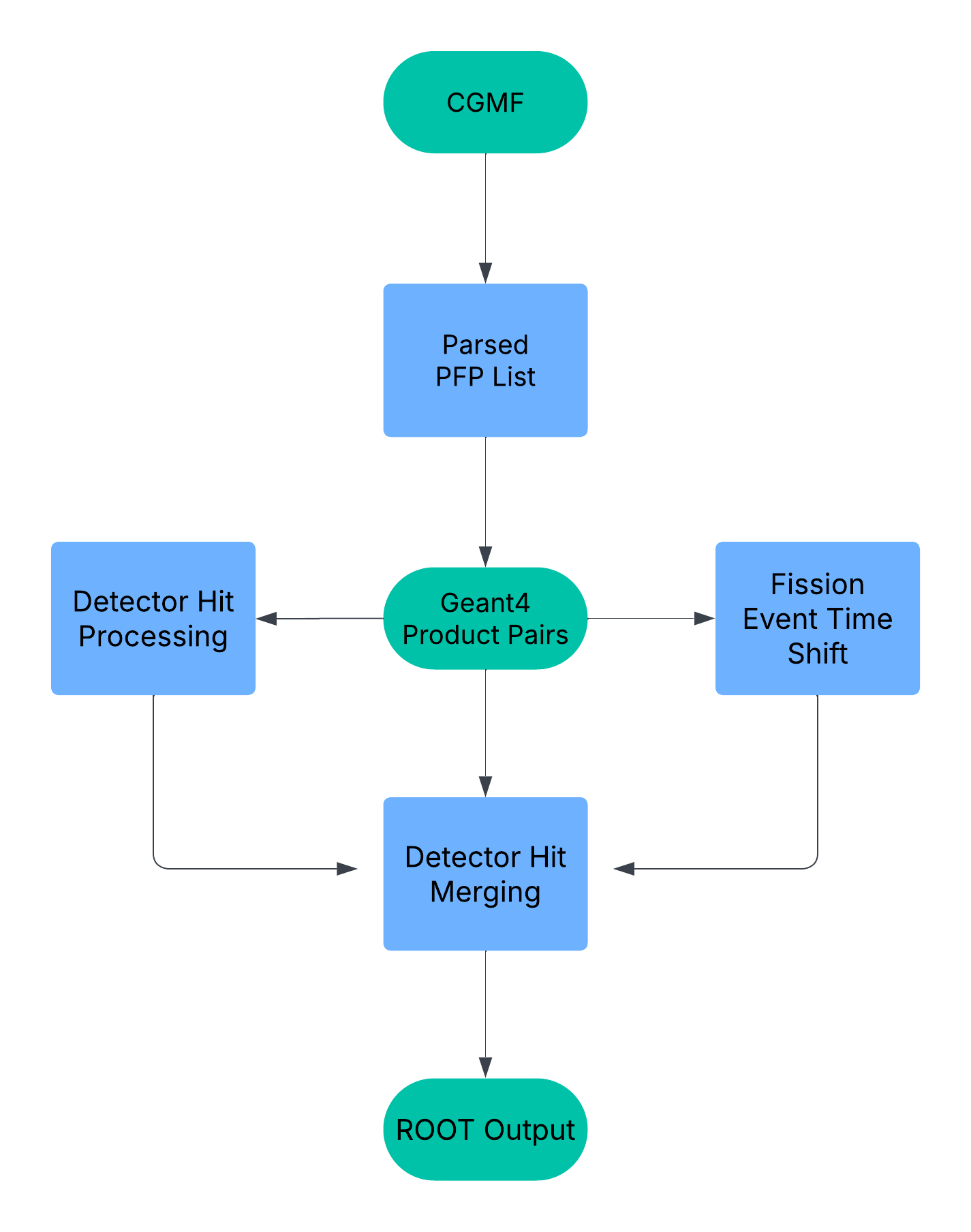}

\caption{Flowchart demonstrating how JANGOFETT receives and processes data. For each pair, each product is simulated as its own event.}\label{Cartoon}
\end{figure}

The process used in JANGOFETT 1.0 is as follows: A CGMF simulation is initiated, generating fission fragments ("FF", pre-neutron evaporation) and primary fission products ("PFP", post-neutron evaporation but before beta decays) per the literature definitions \cite{madland_energy_release_2006}. These are parsed from CGMF outputs into a .csv file which can be alternately used from a prior CGMF run for faster simulations. Geant4 then iterates through the fission event list, and simulates each PFP on a particle-by-particle basis, with separate PFPs from the same fission event in successive simulations. Detector hits are time shifted based on a randomly-sampled Poisson distribution, with the input of an assumed fission rate. Radiation from individual PFPs in the same fission event are time-shifted by the same amount to produce an "experimental time" for correlations. The time-dependent detector responses are then checked for overlapping energy deposition steps within the same detector volume and time windows, then summed together to produce corvariant hits at the end of the Geant4 run. A rolling coincidence window is implemented to associate detector hits with one another. False coincidences occur long after the prompt event that generated the PFPs, but are otherwise indistinguishable from true coincidences. Further detail regarding this is described in a flowchart in Figure \ref{Cartoon}, as well as in the sections below. 

\subsection{Fission Simulation and Parsing}
 CGMF simulates FF de-excitation and prompt decays immediately after nuclear scission. Readers are referred to the literature for details beyond the scope of JANGOFETT \cite{talou_fission_2021}. CGMF is currently used to create the simulated FFs. The list-based output of individual nuclei with properties provided before and after neutron evaporation makes it an ideal source of individual FFs. At this time, the Hauser-Feschbach de-excitation radiation simulated are truncated before injection into Geant4, and the built-in Geant4 physics packages are used for de-excitation from a highly-excited nuclear state.

The CGMF output is then parsed to a .csv file with information on the fragment after it has undergone neutron evaporation. CGMF calculates much of this itself, and the PFP momentum vectors, kinetic energy, nuclear charge, mass, neutron multiplicity, and neutron energy are immediately processed into a new file containing the relevant information. This file is read by Geant4 and used as a list of particles to generate in its simulated environment. The final excitation energy of the individual PFPs are also required, which is calculated as: 

\begin{equation}
    \begin{array}{c}
        EX_{PFP} = KE_I+EX_I+[(A_{FF}\times m_{amu})+\Delta_{FF}] \\
        \quad -(N\times m_n)-[(A_{PFP}-N)(m_{amu})+\Delta_{PFP}]-KE_F-\Sigma KE_n  
    \end{array}
\end{equation}

Here $EX$ refers to the excitation energy. $KE_I$ is the initial kinetic energy before neutron evaporation of the FF, and $KE_F$ is the final kinetic energy afterwards of the PFP. $A_{FF,PFP}$ are the PFP nuclear mass numbers, $m_{amu,n}$ are the mass of 1 amu and 1 neutron, $\Delta_{FF,PFP}$ are the mass defects for the ground state of the nucleus of the outgoing FF and the PFP, N is the number of evaporated neutrons, and $KE_n$ represents the kinetic energies of the evaporated neutrons. The mass defects were retrieved from a repository \cite{huang_ame_2021}. The emitted neutrons created in CGMF are not included in the Geant4 simulation with JANGOFETT 1.0, although their momentum transfer to the PFPs is incorporated. The next update of JANGOFETT 1.01 includes the introduction of these neutrons. It should be noted that one nucleus pair ($^{139}$Sb  with one neutron evaporation from a FF of $^{140}$Sb and $^{112}$Ag) resulted in negative excitation energy by this method, with an average of -0.32 MeV. This error occurred 58 times in $8.7\times10^{7}$ fission events, but 100\% of the time the PFP pair occurred, the error arose. The parser code forced the negative excitation energies to zero in this case. We attribute this to incompatible mass defects between CGMF and \cite{huang_ame_2021}. 

\subsection{Geant4 Simulation}
In Geant4, the G4ParticleGun particle generator fires individual excited PFPs into the world volume from the origin by default. The world volume is the space within the simulations where particles may be tracked and interact with materials, including radiation detector volumes. Most often in Geant4, the particle gun is given a single particle (eg: a nucleus, nucleon, or a fundamental particle) and the desired number of simulations that the particle should be fired and tracked for. Geant4 physics processes incorporate statistical de-excitation models, allowing excited PFPs to be used as primary particles. JANGOFETT uses the G4ParticleGun in such a way that pairs of nuclei (read from the pre-parsed .csv file) are fired for every event in the run. These PFPs are labeled with a fission event ID to facilitate time shifting and debugging. 

As each particle is fired, Geant4 begins a timeline of energy deposition events at time $t = 0$ $(t_0)$, with each following ‘step’ of the particle or its secondaries associated with a time referenced to this moment $t > t_0$. The fired PFP and subsequent radiation proceed to interact with the world as defined by the user-defined geometry. After the particles interact with the world, they will continue to decay to stability, or until they reach a defined time limit. By default, the maximum particle lifetime of $10^6$ seconds (ca. 12 days) has been set due to our specific interests in prompt emissions and lack of interest in long-lived daughters, but this setting can be changed to decay until stability if desired. 

A geometry file, a plain text descriptor containing information on the materials and structures within the Geant4 world volume, is expected for detector definitions, though modifications can be made for those familiar with Geant4. Simulations and analyses presented here use a detector apparatus with a pair of silicon charged-particle detectors in a mocked-up vacuum chamber surrounded by a thin polyimide film. The vacuum chamber is surrounded by high-purity germanium (HPGe) and bismuth germanate (BGO) veto detectors for $\gamma$-spectroscopy of varied geometries. This experimental apparatus is a useful simulation tool for correlating different radiation produced in fission and demonstrating that radiations are produced by Geant4 in a manner that enable multiple events to be associated. 

\subsection{Detector Hit Processing}

In standard Geant4, the time stamp for particles created will always begin at $t_0$. This is not useful for simulations where false coincidences between fission events and their decays are expected at a meaningful rate. At the end of each event, the fission event ID of the particle is considered. Before Geant4 is called by the JANGOFETT script, a list of time shifts is created using a Poisson distribution based on the expected fission rate and number of fission events in the simulation. These time shifts increase sequentially from an overall experimental start time and represent the relative time of a fission event in the simulated experiment. This list is accessed by Geant4 and time shifts are applied to their fission event, creating an experimental time axis.

Once time shifted, the individual steps are summed into hits if they are within a modest time window (20 ns in the default silcon detector implementation). Based on the time and energy resolutions of the detectors recording a given energy deposition, a randomly sampled Gaussian blur is applied to the time and energy values. At the end of the Geant4 run, the valid non-vetoed hits are analyzed together. Hits occurring within the coincidence time window of each other within the same detector volume are summed together, mimicking the response of detectors to multiple incident radiation. These finalized hits are then saved to a structured ROOT output file, which saves the energy deposition, detector volume, and the experimental time of the detection.  

The JANGOFETT 1.0 output that is presented for verification contains only this information, analyzed by a C++ CERN-ROOT code \cite{rene_brun_2019_3895860}. This decision was motivated by the internal use-case for JANGOFETT as it emulates the data structure from fission experiments, which output detector number, energy deposition, and time information for analysis. 

\section{Verification}
JANGOFETT has been benchmarked to ensure that the results it provides are consistent with input CGMF values. A test geometry was used comprising of 21 HPGe detectors with BGO active Compton vetos. This is visualized in Supplementary Figure S.1. Simulations presented here used a $^{252}$Cf spontaneous fission source, modeled with a fission rate of 1000/second and just over five 24 hour days of simulated experimental time.

Due to current computational limitations, the test run described in later sections is simulated with low events. This will be easily managed with the upcoming Collaborative Innovation Complex to be built at Oregon State University. Much larger datasets may be simulated using the new supercomputer, which will provide statistics beyond what is currently reported. The statistics here are sufficient to benchmark the program.

\subsection{Verification Dataset}

\begin{figure}[ht]
\centering
\includegraphics[width=0.9\textwidth]{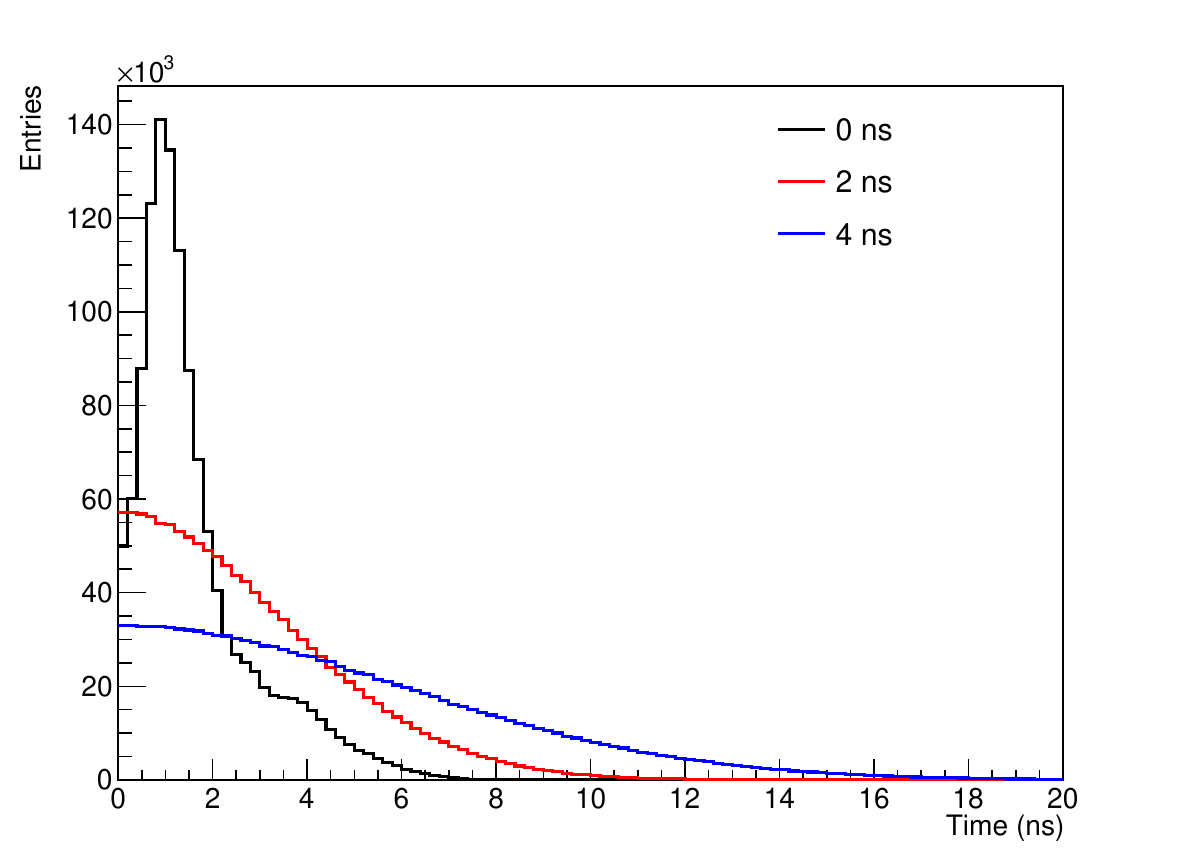}

\caption{Time between coincident PFP energy deposits within silicon detectors for three silicon detector time resolution values (sigma) with a randomly sampled gaussian blur. These figures are generated from a subset of the data for illustration. The time resolution sigma (or FWHM, optionally) is a user input value that can be modified. By default, the silicon detectors have a time resolution sigma of 2 ns, a deliberately conservative estimate for these detectors. This resolution for silicon detectors has been used in all analysis in this work.}\label{FFTime}
\end{figure}

The test setup consisted of two 100 cm$^2$ silicon detectors for measuring total kinetic energy (TKE) of PFPs surrounded by 21 HPGe detectors of varying geometries $\sim$20 cm from the source. Annular BGO detectors surround the HPGes for Compton suppression, which has been employed in the analysis here. 

Coincidence times between PFPs in silicon detectors are shown in Figure \ref{FFTime} and the coincidences between PFPs in the  silicon detectors and gammas are shown in Supplemental Figure S.2. Inspection of Figure \ref{FFTime} indicates a long tail due to time resolution [] that ends before 20 ns. The average time for the two silicon detector signals was chosen as $\Delta_{time}=0$, and 50 ns coincidence windows (-30 ns to +20 ns) were chosen for $\gamma$-PFP correlations.
\FloatBarrier

\subsection{Kinetic Energy Response}
Once coincidence windows were established, TKE was determined by summing the kinetic energies of coincident PFPs. The resulting TKE distribution was fitted with a Gaussian, peaking at 180.083$\pm$0.001 MeV. Individual kinetic energy (KE) spectra were fitted with two summed Gaussians, with peaks observed at 79.544$\pm$0.002  and 102.229$\pm$0.001 MeV, which is within a reasonable calculational uncertainty of existing measurements for $^{252}$Cf(sf) as a calibration standard (181.03, 78.42, and 102.61 MeV) \cite{weissenberger_energy_1986}. These kinetic energies align with the inputs produced by CGMF, indicating no bias in our analysis and steps-to-hits processing, other than being 0.004(4) MeV lower than inputs due to energy losses by detector X-rays and delta electrons. This is an insignificant difference (see Figure S.4).

\begin{figure}[ht]
    \centering
    \subfloat[Kinetic Energy from coincident PFP energy depositions in silicon detectors.]{\includegraphics[width=0.9\textwidth]{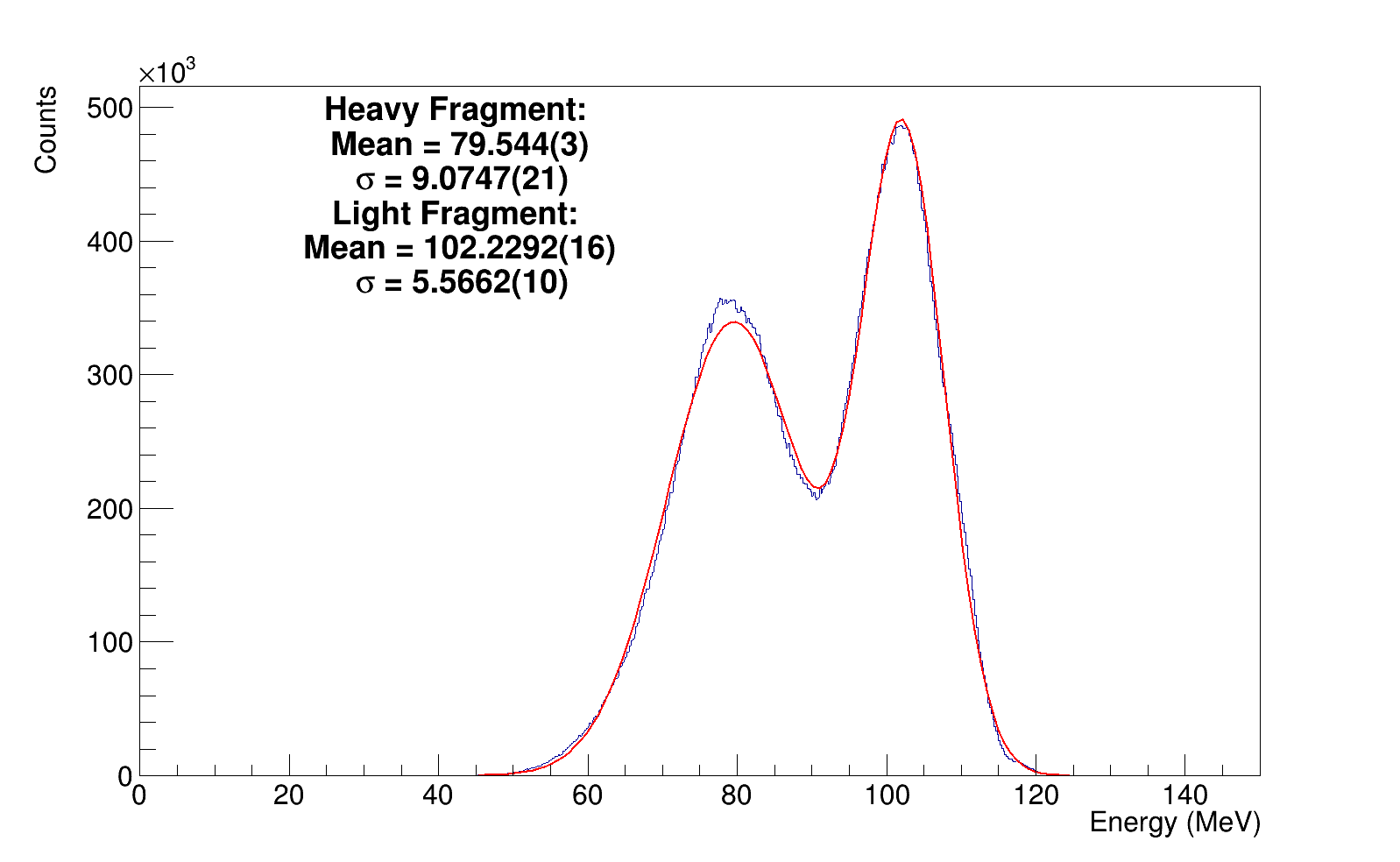}}
    
    \subfloat[Total Kinetic Energy from sum of coincident PFP energy depositions in silicon detectors.]{\includegraphics[width=0.9\textwidth]{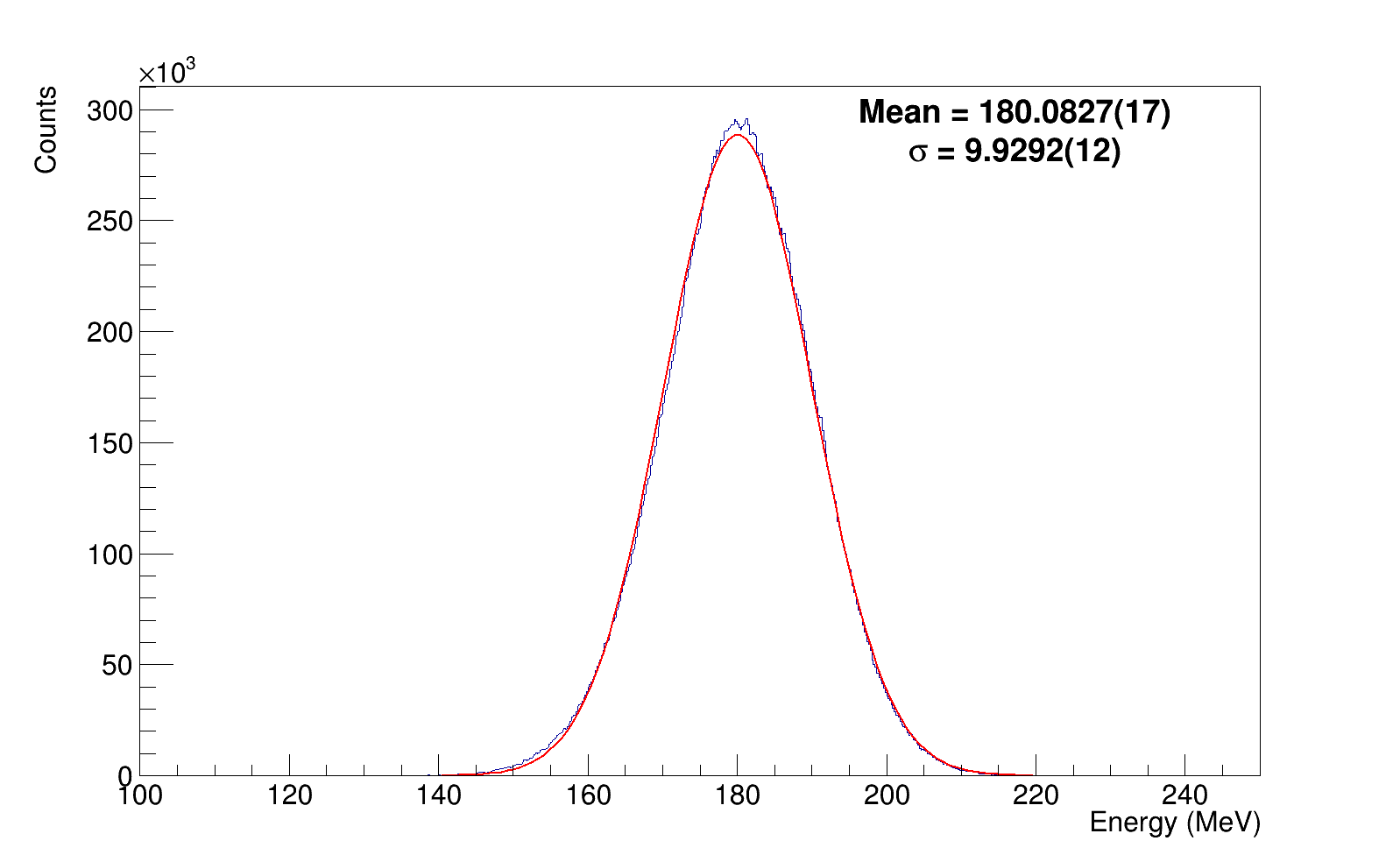}}
  
    \caption{KE (a) and TKE (b) for the dataset with Gaussian fits in red. KE was determined by the energy in each silicon detector, considering fragments coincident within 20 ns. These KE were summed to produce TKE for coincident events.} 

    \label{fig:KE_TKE}
\end{figure}

Figure \ref{fig:KE_TKE} shows a pair of KE (individual PFPs) and TKE (summed PFP energy) plots. Silicon detector signals with less than 20 MeV were excluded. As can be seen in Supplementary Figure S.3, there are a number of TKE signals that occur at $\sim$90 MeV and $\sim$180 MeV above the TKE peak. These are from false  silicon detector triple and quadruple coincidences that are present at a physically meaningful rate when considering the coincidence window. The TKE was analyzed before and after Geant4 simulations to rule out systematic errors - see the Supplementary Material for more detail (Figure S.4).

\begin{figure}[ht!]
\centering
\includegraphics[width=1\textwidth]{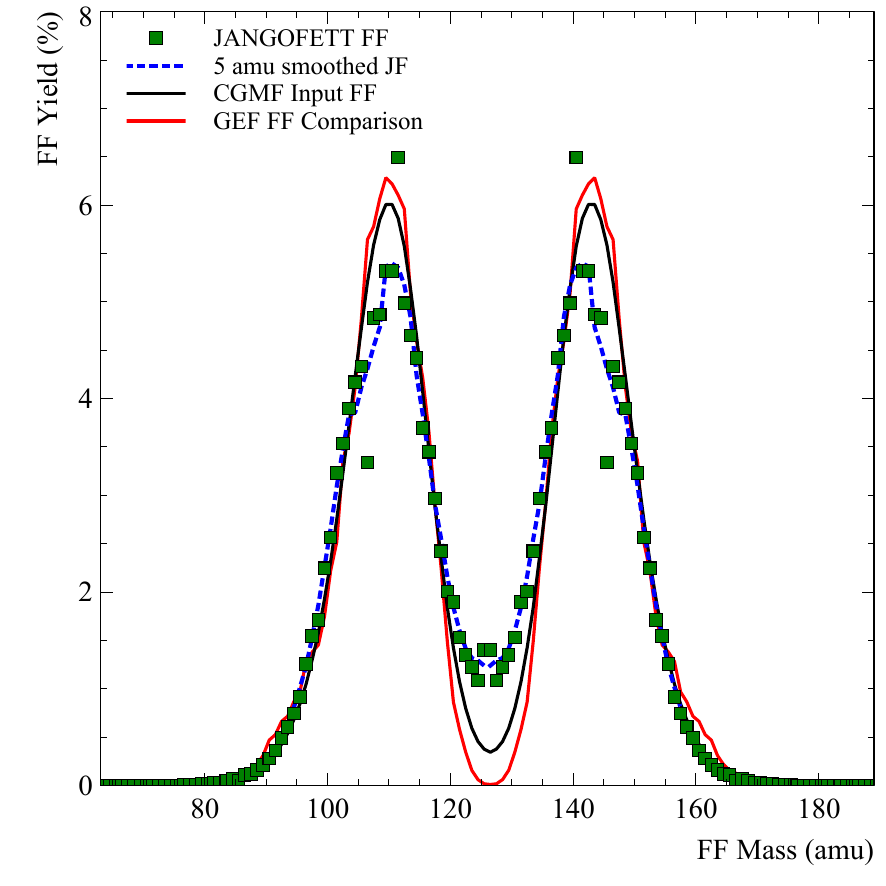}

\caption{Comparison of inferred FF mass distributions: (1) JANGOFETT 2E-method estimates based on KEs, without unfolding or further post-processing; (2) the input FF mass distribution from CGMF\cite{talou_fission_2021}; and (3) comparative predictions from the GEF code\cite{schmidt_gef_2016}. A version of the JANGOFETT distribution smoothed with a 5 amu-wide binning is also shown to illustrate overall yield trends and approximate experimental mass resolution\cite{wagemans_fission_1991}. Note the lower peaks and broader symmetric fission branch due to 2E method broadening, compared to the input from CGMF.}\label{Mass}
\end{figure}

The PFP KE was further used to approximate PFP mass via the double-energy (2E) method \cite{schmitt_fragment_1966}, yielding mass distribution with a precision of approximately $\pm$4 atomic mass units (amu). The pre-neutron evaporation FFs are then reconstructed via calculations of $^{252}$Cf(sf) with neutron multiplicity as a function of post-neutron evaporation from GEF\cite{schmidt_gef_2016}. This method has been utilized in similar mass yield measurements in recent articles \cite{king_tke_2017}\cite{chemey_pu239_2020}\cite{pica_pu240_2022}. Figure \ref{Mass} demonstrates the 2E-method mass yields without any additional gating conditions beyond timing coincidence and compares estimated yields to input FF masses from CGMF in addition to yields produced from GEF.

\subsection{\texorpdfstring{$^{252}$Cf(sf)}{252Cf} PFP Identification}

With these coincidence criteria established, specific PFP pairs were selected for verification studies. Four heavy PFPs were selected for gating, $^{141}$Cs, $^{144}$Ba, $^{150}$Ce $^{149}$Nd.$^{144}$Ba is used for analysis in this work, while the remainder are included in the Supplementary Material. To isolate specific PFP events, an initial mass gate was applied, selecting events within 4 amu of the desired isotope, based on the 2E mass estimate. A second coincidence gate was then imposed, requiring a $\gamma$ coincidences from the heavy PFP to be measured in the HPGe detectors without a BGO Compton veto. $\gamma$-rays for coincidence were selected from a subset of high-probability $\gamma$-rays in the established nuclear data. Finally, an additional $\gamma$-gate was applied for each of the light PFP possible with 0-5 neutrons evaporated from the pair. In this analysis, only information that could be extracted from experimental methods was used to determine the PFP pair. As a result, the 0-5n values represent neutrons removed from the system agnostic of which FF they were removed from. For these selected PFPs, Coincident $\gamma$ histograms were generated to demonstrate both heavy and light PFP were simultaneously detected.

All peaks in the histograms were evaluated to determine whether the identified fragments could be plausibly misidentified by mass and $\gamma$-gates and should be discarded. Our evaluations did not determine that any fragments needed to be discarded in this way, as all $\gamma$ signals that were above background were able to be identified with reasonable certainty. The results from the  $\gamma$ singles plots when gated on 2E-masses are shown in Figure 5 (see Figure S.5 - S.7 for other gated fragments and Table S.1-S.4 for a list of the $\gamma$ energies used to gate in the Supplemental Material).  

\label{Results}
\begin{figure}[ht]
    \centering
    \vspace{-22pt}
    \subfloat[1n]{\includegraphics[width=0.5\textwidth]{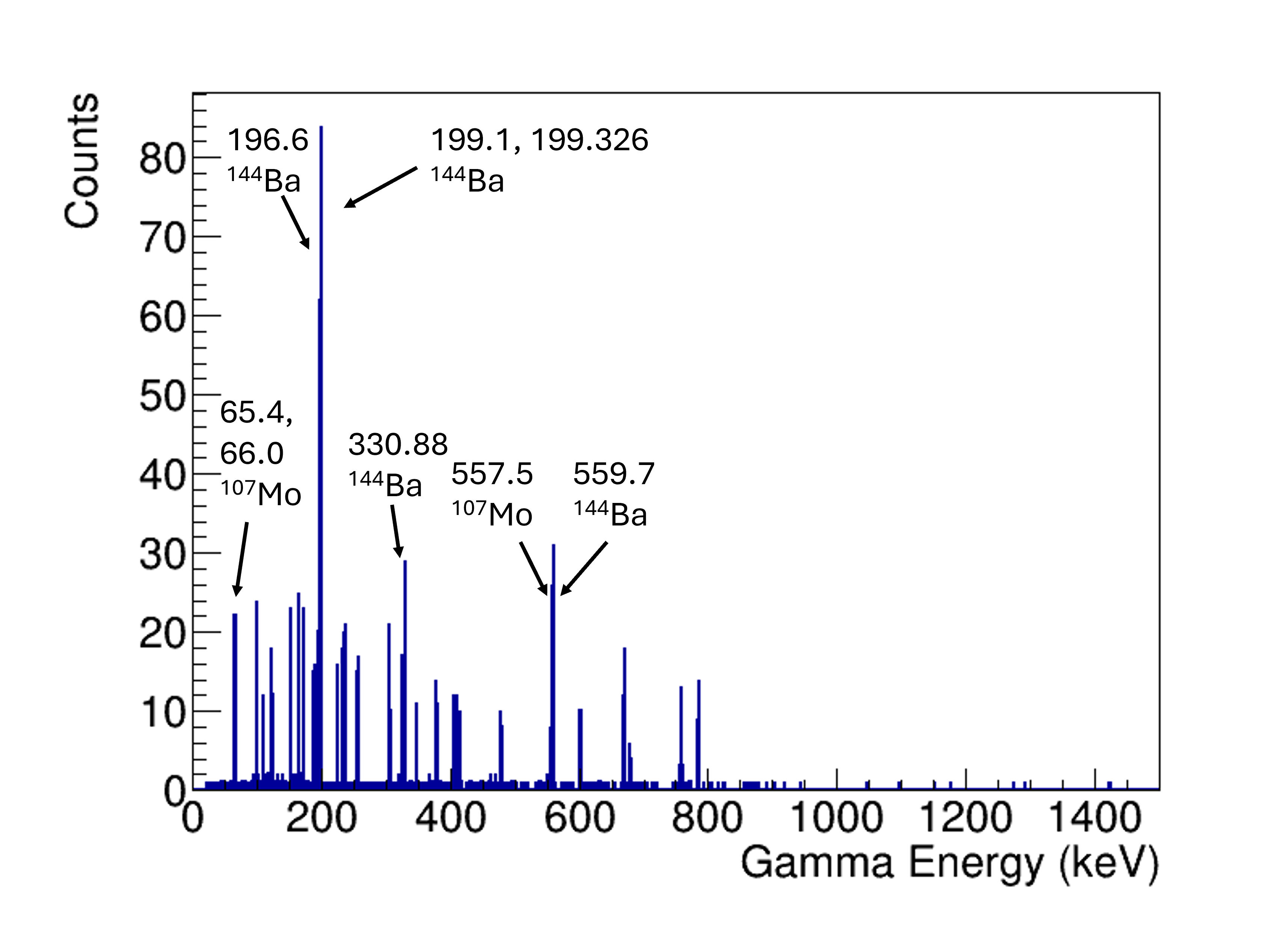}}
    \subfloat[2n]{\includegraphics[width=0.5\textwidth]{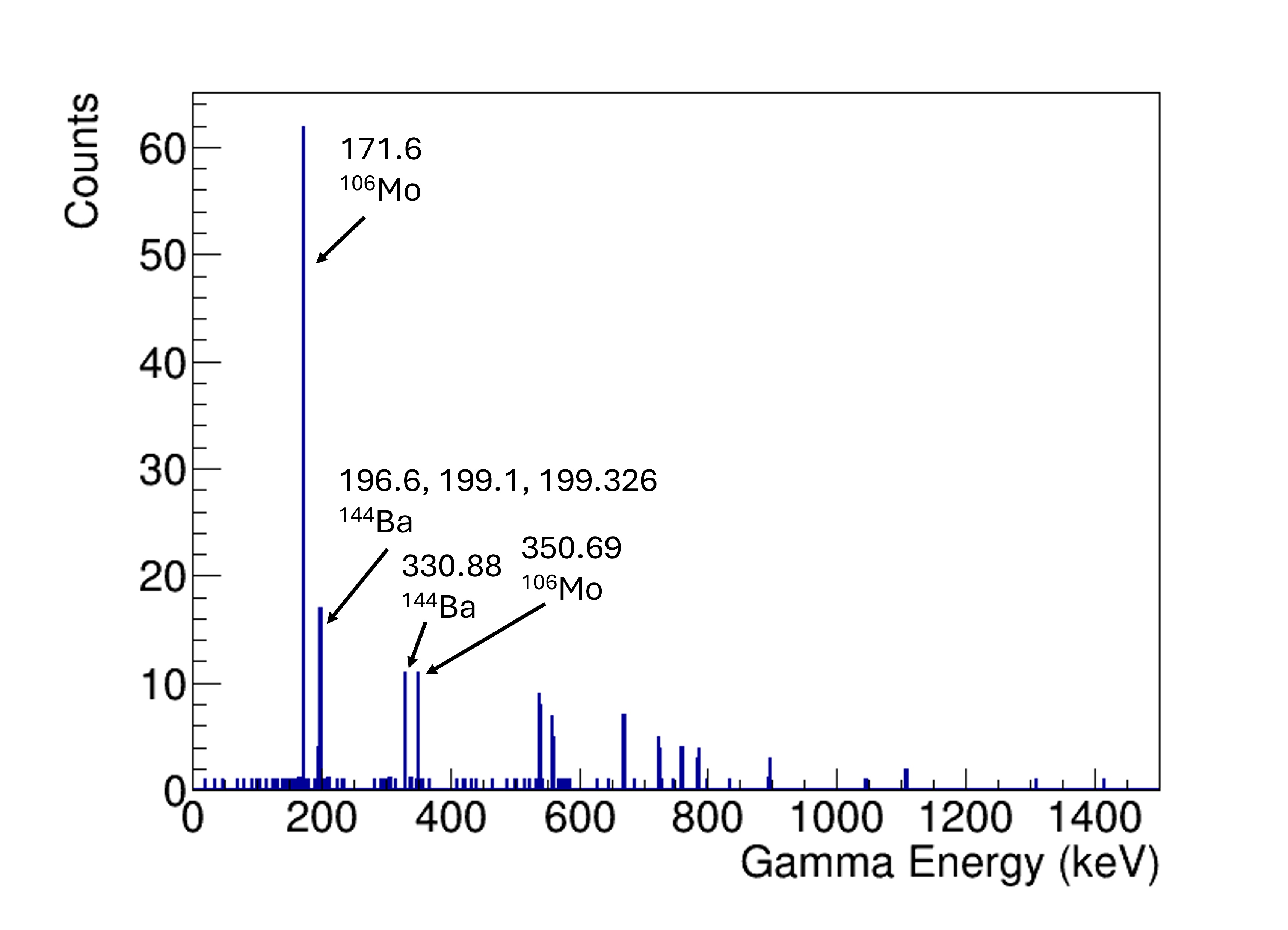}}
    \vspace{-10pt}
    
    \subfloat[3n]{\includegraphics[width=0.5\textwidth]{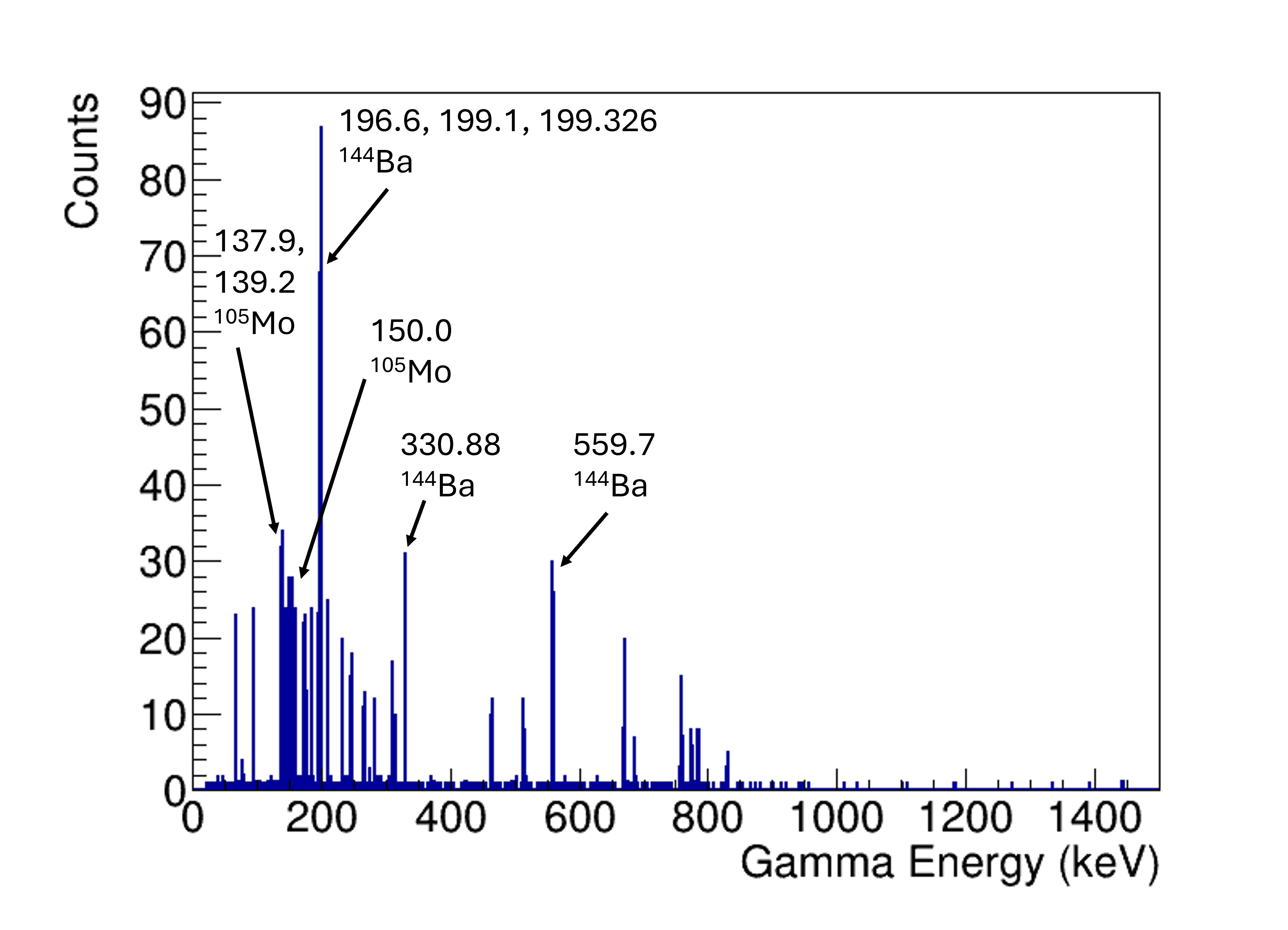}}
    \subfloat[4n]{\includegraphics[width=0.5\textwidth]{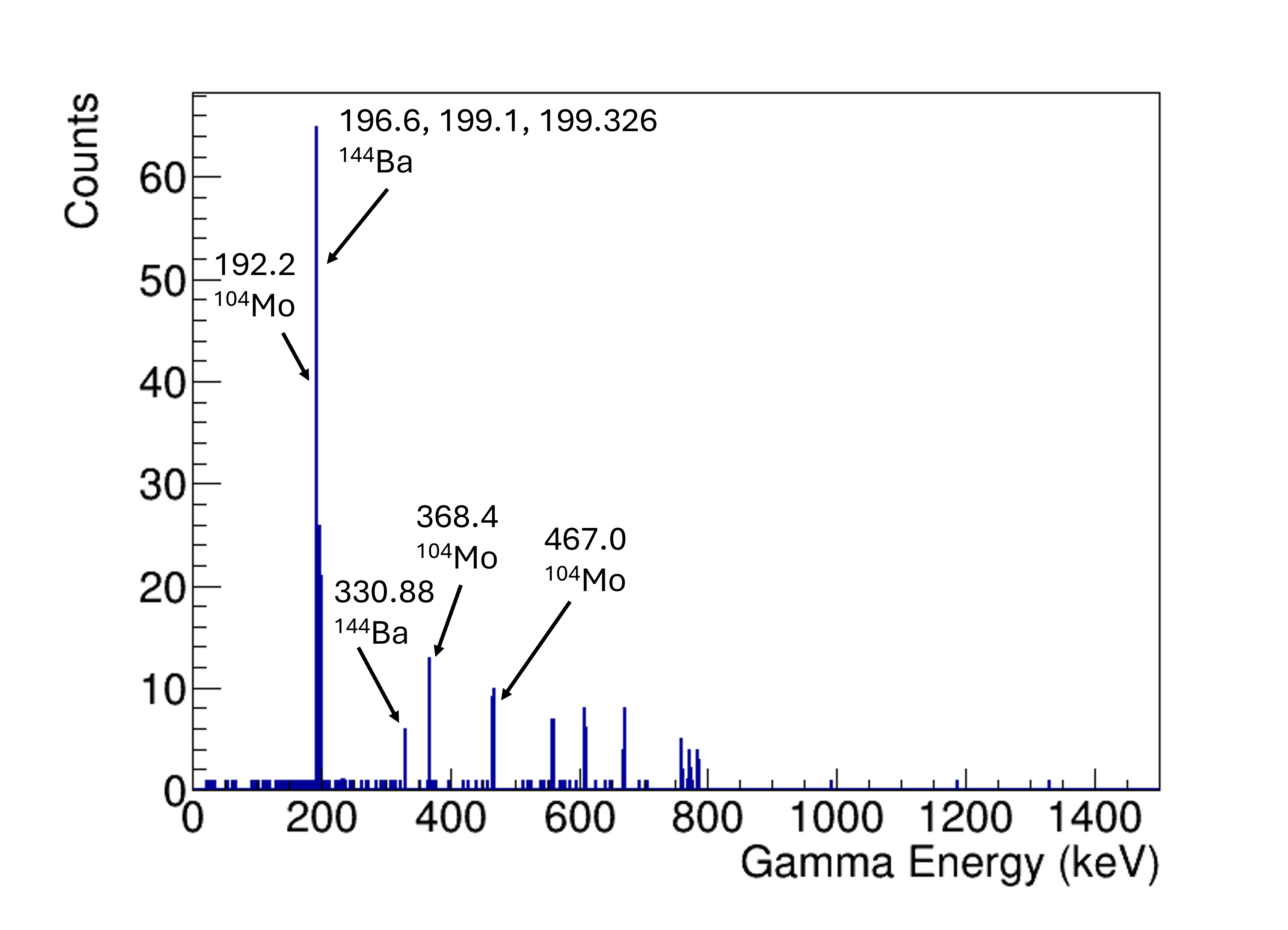}}
    \vspace{-10pt}
    
    \subfloat[5n]{\includegraphics[width=0.5\textwidth]{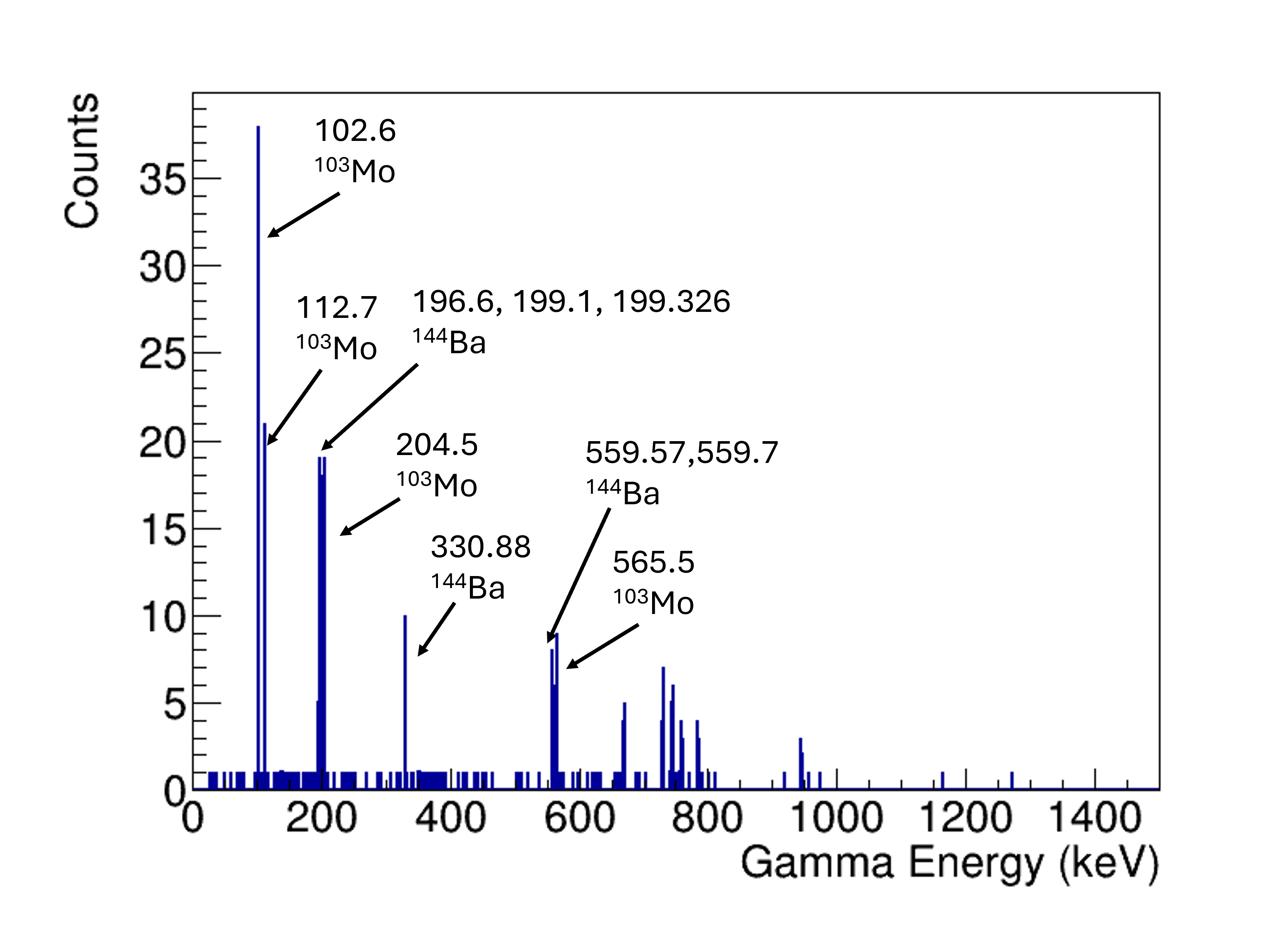}}
    \subfloat[6n]{\includegraphics[width=0.5\textwidth]{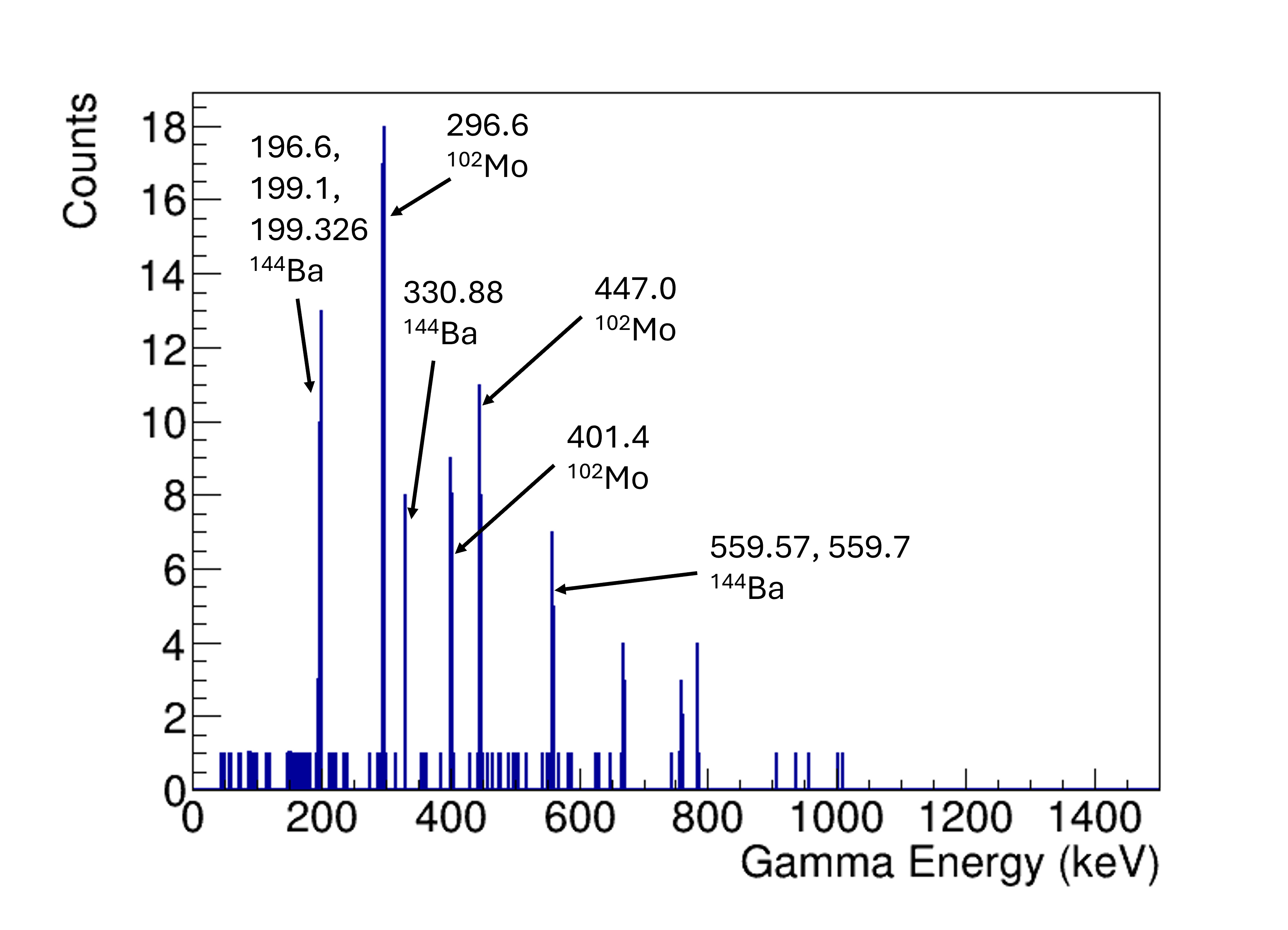}}

    \caption{$\gamma$ energy spectra when at least one coincident $\gamma$ from $^{144}$Ba and at least one $\gamma$ from the correlated PFP (i.e. $^{252}Cf\rightarrow ^{144}Ba+^{108-x}Mo+xn$), for $x$ between 1 and 6 for (a) through (f). Energy labels on the figure are given in keV with the respective PFP label.}
    \label{fig:ggg}
\end{figure}

\begin{figure}[ht]
    \centering
    \vspace{-22pt}
    \subfloat[1n]{\includegraphics[width=0.5\textwidth]{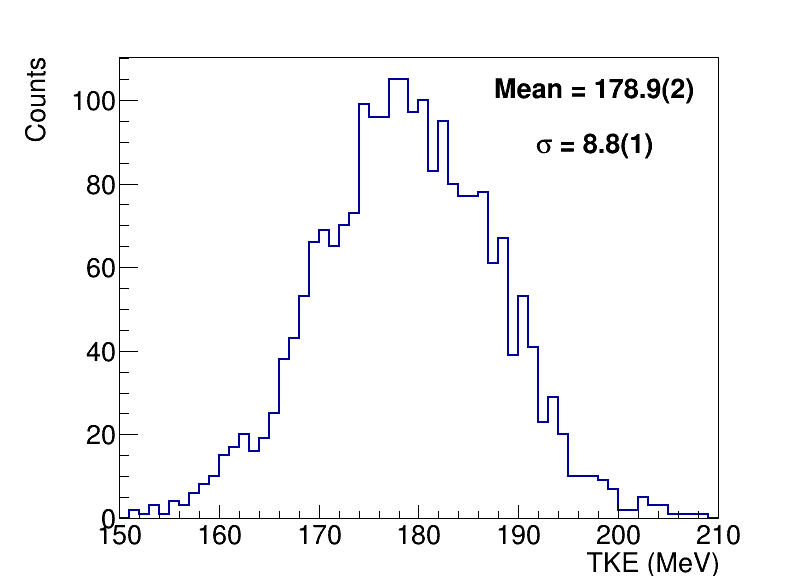}}
    \subfloat[2n]{\includegraphics[width=0.5\textwidth]{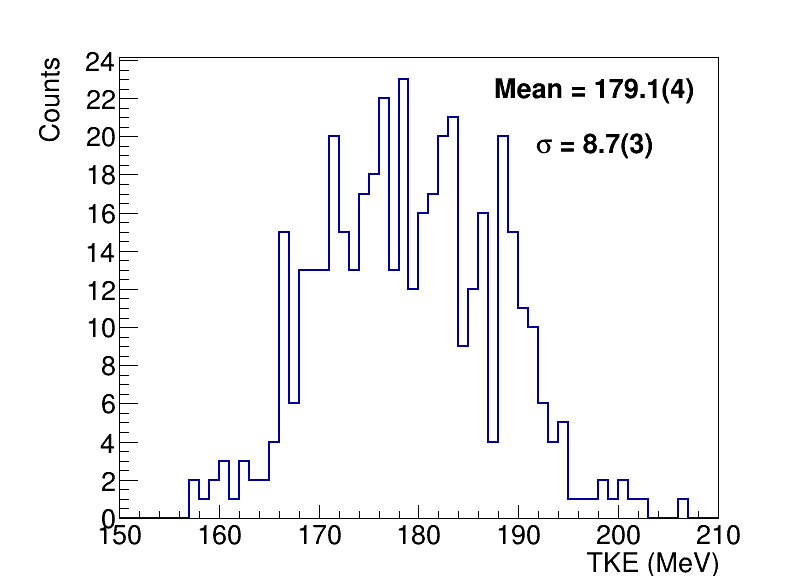}}
    \vspace{-10pt}
    
    \subfloat[3n]{\includegraphics[width=0.5\textwidth]{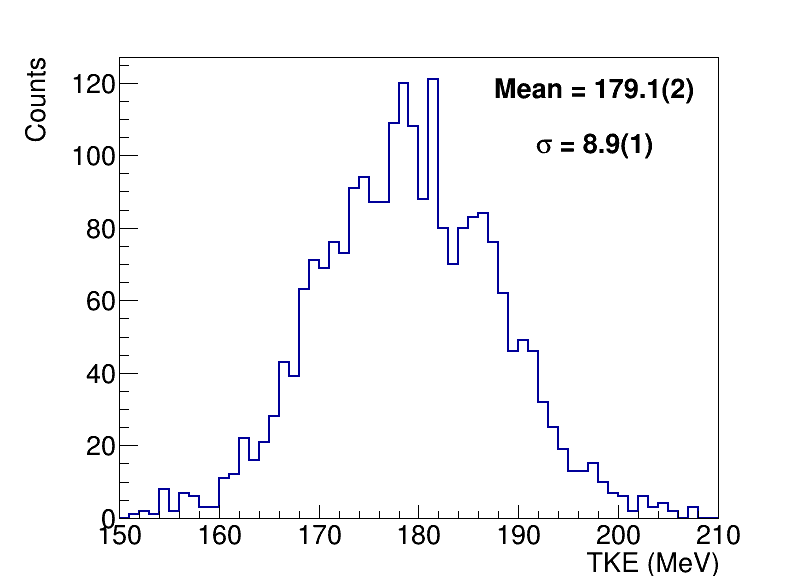}}
    \subfloat[4n]{\includegraphics[width=0.5\textwidth]{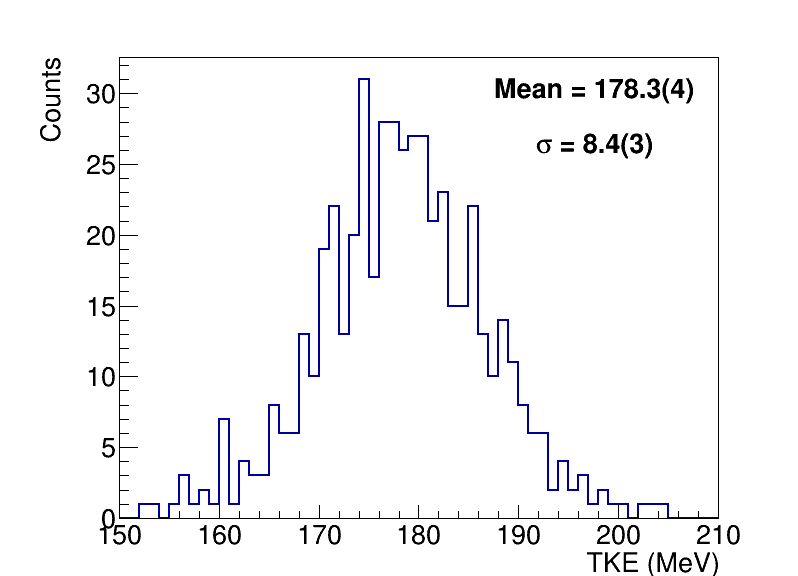}}
    \vspace{-10pt}
    
    \subfloat[5n]{\includegraphics[width=0.5\textwidth]{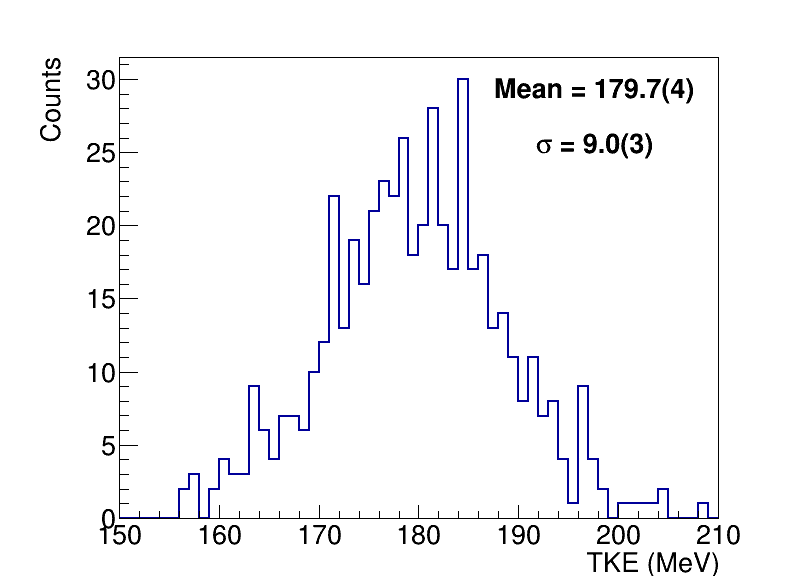}}
    \subfloat[6n]{\includegraphics[width=0.5\textwidth]{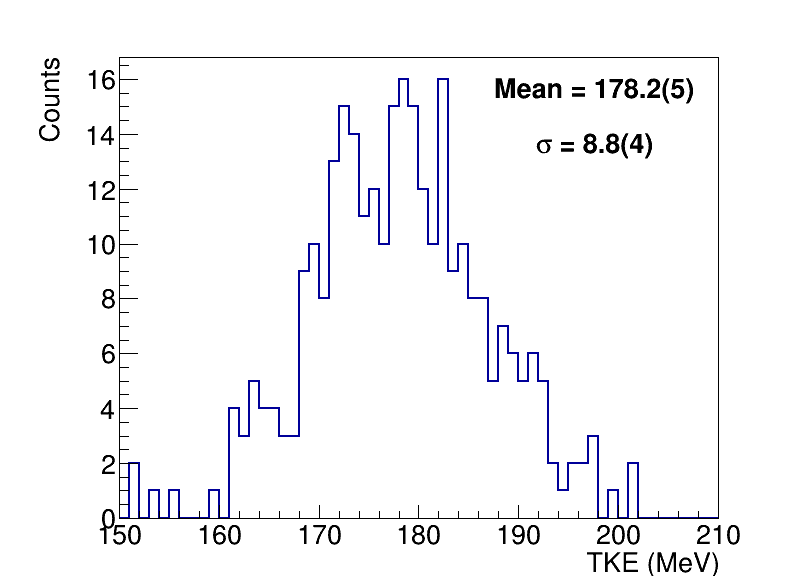}}

    \caption{TKE plots for tagged 144Ba + Light PFP events from at least one $\gamma$ from 144Ba and at least one $\gamma$ from the correlated light PFP. It should be noted that expected trends between TKE and TXE are not observed due to low statistics and overgeneralized $\gamma$ gates resulting in the inclusion of events not attributed to 144Ba + its Light PFP. This is expected to be improved with an increase in simulated fission events and more conservative gating requirements. This figure is included to demonstrate the ability to isolate multiple particle events using time correlations in Geant4 with JANGOFETT. }
    \label{fig:TKE_0-6n}
\end{figure}

After identifying fragments from mass estimates and coincident $\gamma$, the KE of each fragment was summed to produced TKE plots for each of the light fragments observed, which are displayed in Figure \ref{fig:TKE_0-6n}. It is difficult to conclude much after only 24 hours of simulated data, but the presented figures are sufficient to conclude that multiple fragments are simulated within a single time-correlated event, distinguishable from random background.

\FloatBarrier
\section{Summary}
JANGOFETT is a new tool for Geant4, developed to read primary fission product pairs into Geant4 with simulated time-based correlations. JANGOFETT performance was verified using a simulation of $^{252}$Cf(sf) detector responses. Verification that the simulation produces distinguishable list-mode time sorted hits in detector volumes was performed by isolating coincidences from specific heavy and light PFPs, based on mass estimates and $\gamma$ energy. A heavy PFP of $^{144}$Ba and its 0-5n light complements are shown here, with $\gamma$ signals for each partner PFP appearing in the same simulation time, though they came from individual Geant4 instances. Additional PFPs were analyzed and are included in the Supplementary Material. This tool enables the simulation and analysis of covariant fission observables in Geant4, generating data structures that are similar to those produced by experimental apparatus in fission for analysis.

JANGOFETT 1.0 has been made available at \url{https://github.com/AlchemeyLab/JANGOFETT} under GNU General Public License v3.0.

\section*{Acknowledgments}

Authors LW and JS acknowledge fellowships from the Nuclear Regulatory Commission University Nuclear Leadership Program administered through Oregon State University. Authors gratefully acknowledge a startup package from the Oregon State University College of Engineering. Author LW acknowledges additional support from the ARCS foundation Oregon.

\section*{CRediT authorship contribution statement}
\textbf{L. Walker:} Methodology, Software, validation, writing - original draft, writing - review and editing, visualization. 
\textbf{J. Shire:} Software, validation, writing - original draft, writing - review and editing, visualization. 
\textbf{J. Jaffe:} Software, validation, Writing - review and editing, visualization. 
\textbf{P. Sprando:} Software, validation, Writing - review and editing, visualization. 
\textbf{J. Olinger:} Writing - review and editing, visualization.
\textbf{A. Chemey:} Conceptualization, methodology, software, writing - original draft, writing - review and editing, supervision, project administration.

\end{document}